\documentclass[a4paper,12pt]{article}
\usepackage{graphics}
\newcommand{\refeq}[1]{(\ref{#1})}
\newcommand{\pk}[1]{{\rm #1}}
\newcommand{\bmath}[1]{\mbox{\boldmath ${#1}$}}
\newcommand{\beps}{\bmath{\epsilon}}
\newcommand{\ip}[2]{\ensuremath{{#1}\!\cdot\!{#2}}}
\newcommand{\dd}{{\rm d}}
\newcommand{\dkub}[1]{{\rm d}^{3}\!{#1}}

\begin{document}
\setlength{\unitlength}{1mm}

\begin{titlepage}
\vspace{2.0cm}

\begin{center}
{\LARGE\bf An explanation of the ABC} \\[2ex]
{\LARGE\bf enhancement in the dd~\mbox{\boldmath\LARGE $\rightarrow\alpha$}X}
\\[3ex]
{\LARGE\bf reaction at intermediate energies}\\[8ex]

Anders G{\aa}rdestig\footnote{Electronic address grdstg@tsl.uu.se} and
G\"oran F\"aldt\footnote{Electronic address faldt@tsl.uu.se}\\[1ex]
Division of Nuclear Physics, Uppsala University, Box 535,\\
S-751 21 Uppsala, Sweden\\[4ex]
Colin Wilkin\footnote{Electronic address cw@hep.ucl.ac.uk}\\[1ex]
University College London, London WC1E 6BT, UK\\[8ex]
\end{center}

\baselineskip 3ex
\begin{abstract}
The \mbox{dd $\rightarrow\alpha$X} reaction is studied in a model where
each pair of nucleons in the projectile and target deuterons undergoes pion
production through the \mbox{NN $\rightarrow \pk{d}\pi$} reaction. The 
condition that the two deuterons fuse to form an $\alpha$-particle then leads to
peaks at small missing masses, the well-known ABC enhancement, but also a broad
structure around the maximum missing mass. With a simplified input amplitude the
model gives a quantitative description of both the $\alpha$-particle momentum
and angular distributions for a deuteron beam energy of 1250~MeV.
\end{abstract}
\vspace{1cm}

\centerline{\today}
\vspace{1cm}


\end{titlepage}

\baselineskip 4ex
Measurements of the $\alpha$-particle momentum spectra in the 
\mbox{dd $\rightarrow\alpha$X} reaction reveal rich structure for beam energies
throughout the $0.8-1.4$~GeV region~\cite{Ban}. In particular there are very 
sharp peaks near the kinematic limits, {\it i.e.}\ the maximum in the
$\alpha$-particle c.m.\ momentum, and these correspond to missing masses of 
$M_{\rm X}\sim\:$300~MeV. In addition, however, there is a broad central bump
around the maximal missing mass, {\it i.e.}\ in the region where the 
$\alpha$-particle is at rest in the c.m.~system. Rather similar structures were 
observed many years ago in both the \mbox{pd 
$\rightarrow\, ^{3}$He\,X}~\cite{ABC} and 
\mbox{np $\rightarrow$ d\,X}~\cite{Homer,Hall} reactions and this low 
missing-mass peak is commonly known as the ABC-effect. It is clear from the 
mass of the ABC that it must consist of two pions and the absence of an 
equivalent peak in the \mbox{pd $\rightarrow\ ^{3}$H\,X$'$} case~\cite{ABC} 
shows that it must have isospin $I=0$. It is therefore likely that the ABC 
manifests itself so clearly in deuteron-deuteron collisions because it leads 
to a pure $I=0$ channel for the pions.

The authors of ref.~\cite{ABC} originally suggested that the ABC might
correspond to a large $s$-wave isoscalar $\pi$-$\pi$ scattering length
but phase shifts obtained from high energy pion production experiments show this
is in fact very small~\cite{Martin}. Since the mass and width of the ABC 
peak vary with kinematic conditions, it is natural to see if it can be 
understood as a kinematic effect associated with the presence of nucleons or 
nuclei. In this spirit Risser and Shuster~\cite{Risser} explored the 
\mbox{np $\rightarrow$ d\,X} reaction in a model where two different 
$\Delta$-resonances were excited through pion exchange. Though they could 
generate peaks in the right regions, the calculations had limited 
quantitative success, but it is not clear whether this was due to the 
simplifications in their formalism, and in particular their neglect of 
spin~\cite{Risser,Barry}. Certainly double $p$-wave pion production should 
lead to kinematic enhancements in the spinless case because the simplest matrix 
element squared is proportional to $(\bmath{k}_1\cdot\bmath{k}_2)^2$, 
{\it i.e.}\ the square of the scalar product of the two pion momenta. This is 
maximal when the two momenta are parallel (the ABC peak) and antiparallel (the 
central bump).

ABC production in deuteron-deuteron scattering is at its largest around 600~MeV
per nucleon~\cite{Ban} and this is close to the maximum of single-pion 
production in \mbox{pp $\rightarrow \pk{d}\pi^+$} arising from 
$\Delta$-excitation. It therefore seems natural to consider two-pion production
as coming from two such independent reactions, involving two different pairs of
nucleons from the projectile and target deuterons, where the $\alpha$-particle 
is formed when the two final deuterons stick together. If we neglect the Fermi 
motion in the initial deuterons, the c.m.~systems of the deuteron-deuteron and 
nucleon-nucleon channels coincide so that the produced deuterons (and pions) 
will have the same c.m.~momenta. The relative momentum in the final d-d system 
will therefore be very small when the pions come out with low excitation energy 
in the ABC peak and there will then be little suppression coming from the 
$\alpha$:dd sticking factor. This classical argument of favourable kinematics 
provides another reason why ABC production might be seen clearest in the 
\mbox{dd $\rightarrow\alpha$X} case.

\begin{figure}[p]

\newcommand{\qism}[1]{\ensuremath{\scriptstyle\bf q}_{\scriptscriptstyle #1}}
\begin{picture}(120,60)(-40,-10)


    \put(-15,42){\makebox(0,0)[b]{$\pk{d_{2}}$}}
    \put(-25,41){\line(1,0){20.101}}
    \put(-25,39){\line(1,0){20.101}}
    \put(-15,38){\makebox(0,0)[t]{$\bmath{p}$}}

    \put(0,40){\circle{10}}
    
    \qbezier(7,39)(8,36)(4.95,35.05)
    \put(8,36){\makebox(0,0)[tl]{$\bmath{q}_{2}$}}
    \put(5,40){\line(1,0){20}}

    \put(30,40){\circle{10}}

    \put(50,45){\makebox(0,0)[br]{$\bmath{k}_{2}$}}
    \multiput(34.851,41.213)(5,1){6}{\line(5,1){3.7}}


    \put(-15,2){\makebox(0,0)[b]{$\pk{d_{1}}$}}
    \put(-25,1){\line(1,0){20.101}}
    \put(-25,-1){\line(1,0){20.101}}
    \put(-17,-2){\makebox(0,0)[t]{$-\bmath{p}$}}

    \put(0,0){\circle{10}}

    \qbezier(7,1)(8,4)(4.95,4.95)
    \put(5,0){\line(1,0){20}}
    \put(8,4){\makebox(0,0)[bl]{$\bmath{q}_{1}$}}

    \put(30,0){\circle{10}}

    \multiput(34.851,-1.213)(5,-1){6}{\line(5,-1){3.7}}
    \put(50,-5){\makebox(0,0)[tr]{$\bmath{k}_{1}$}}


    \put(3,4){\line(3,4){24}}

    \put(3,36){\line(3,-4){10}}
    \qbezier(13,22.667)(18,24)(17,17.333)
    \put(27,4){\line(-3,4){10}}

    \put(48,30){\makebox(0,0)[bl]{$\rm d'$}}
    \put(33.526,36.455){\line(3,-2){21.845}}
    \put(34.628,38.108){\line(3,-2){21.845}}

    \put(34.628,1.892){\line(3,2){21.845}}
    \put(33.526,3.545){\line(3,2){21.845}}
    \put(48,11){\makebox(0,0)[tl]{d}}

    \put(60,20){\circle{10}}

    \put(64,17){\line(1,0){21}}
    \put(64.899,19){\line(1,0){20.101}}
    \put(64.899,21){\line(1,0){20.101}}
    \put(64,23){\line(1,0){21}}
    \put(75,16){\makebox(0,0)[t]{$\bmath{p}_{\alpha}$}}

\end{picture}
    \caption{Feynman diagram for the ${\rm dd} \rightarrow
    \alpha\,\pi\pi$ reaction showing the momenta in the c.m.\ system.}
\label{fig:ddapp}
\end{figure}
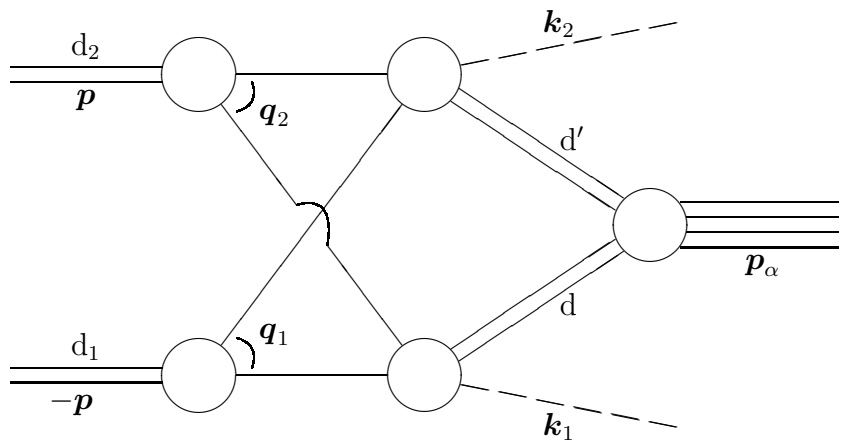

The Feynman diagram corresponding to our model for
\mbox{dd $\rightarrow\alpha\pi^+\pi^-$} is shown in fig.~\ref{fig:ddapp}, 
where the various momenta are defined in the c.m.~system. In the 
non-relativistic limit of the Fermi momenta, the matrix element deduced from 
this diagram is
\begin{eqnarray}
    \mathcal{M} & = &
    \frac{-\sqrt{2\pk{m_{\alpha}}}\,(\ip{\beps_{1}}{\beps_{2}})}
    {\gamma^{2}(2\pi)^{\frac{3}{2}}}
    \int \dkub{q_{1}}\dkub{q_{2}}\,
    \varphi_{\pk{d}}(\bmath{q}_{1}') \varphi_{\pk{d}}(\bmath{q}_{2}') 
    \psi_{\alpha}^{\dagger}(\bmath{q}_{\alpha}) 
    \nonumber \\
    & \times &
    \sum_{\rm spins} \frac{1}{\sqrt{3}} (\ip{\beps}{\beps'})
        M_{\pi}^{1}
        \frac{1}{\Delta E_{1}+i\epsilon}
        M_{\pi}^{2}
    +(\Delta E_{1}\leftrightarrow \Delta E_{2}),
\label{e_1}
\end{eqnarray}
where the $({\beps},\,{\beps}')$ are the polarization vectors of the internal
deuterons and $({\beps}_1,\,{\beps}_2)$ those of the incident. The wave 
functions of the final $\alpha$-particle and the initial deuterons 
$\psi_{\alpha}(\bmath{q}_{\alpha})$ and $\varphi_{\pk{d}}(\bmath{q}_{i}')$
depend upon the Fermi momenta in the rest systems of the corresponding nuclei.
In the deuteron case the longitudinal component is boosted with respect to the 
c.m.~system to give $\bmath{q}_{i}'=(\bmath{q}_{i}^{b},q_{i}^{z}/\gamma)$, 
where $\gamma=E_{\pk{d}}/\pk{m_{d}}$ and $b$ and $z$ denote the transversal and
longitudinal components. The analogous relativistic effect is 
small for the $\alpha$-particle and so $\bmath{q}_{\alpha}=\bmath{q}_{1}-
\bmath{q}_{2}-(\bmath{k}_{1}-\bmath{k}_{2})/2$. Since our model is not dependent
upon exotic Fermi momentum components, it is reasonable to neglect the $D$-state
components in the deuterons and the $\alpha$-particle.

The matrix element in eq.~\refeq{e_1} is essentially that of second order 
perturbation theory, where the energy denominators corresponding to the two
different time orderings are
\begin{equation}
\label{e_2}
    \Delta E_{i} = \mp\,\left\{\ip{\bmath{v}_{\pk{d}}}{(\bmath{q}_{1}
    +\bmath{q}_{2})}+(\omega_1-\omega_2)/2\right\}\:,
\end{equation}
with $\bmath{v}_{\pk{d}}$ being the c.m.\ deuteron velocity and $\omega_{i}$
the pion total energies. This simple result only follows when retaining just
the linear terms in the nuclear Fermi momenta. In such a case the two principal
value integrals in eq.(\ref{e_1}) cancel to leave only the $\delta$-function 
term.

The largest \mbox{pp $\rightarrow \pk{d}\pi^+$} amplitude in the $\Delta$ region
corresponds to the $^{1\!}D_{2}p$ transition~\cite{SAID}, for which
\begin{equation}
\label{e_3}
    M_{\pi}  =  \mathcal{C}
    \left[ 3(\ip{\bmath{\hat{p}}}{\bmath{k}})
    (\ip{\bmath{\hat{p}}}{{\beps}^{\dagger}})
    -\ip{\bmath{k}}{{\beps}^{\dagger}}
    \right],
\end{equation}
where $\bmath{p}$ and $\bmath{k}$ are the proton and pion c.m.\ momenta and
$\mathcal{C}$ is a function of energy. This form does not depend upon the 
proton spin variables and it is due to this that we have the simplification in 
eq.(\ref{e_1}) which leaves only an $\ip{\beps_{1}}{\beps_{2}}$ dependence. 
Keeping only this term, the value of $\mathcal{C}$ is fixed by experimental 
\mbox{pp $\rightarrow \pk{d}\pi^+$} data in the forward direction, as 
summarised in the SAID database~\cite{SAID}.

Some care must be taken over the effect of the $\alpha$-particle binding 
energy in determining the energy in the proton-proton system. At the ABC-peak
the pions emerge with zero relative momentum and we assume that the laboratory
kinetic energy in the $\pi^+$d system in the inverse reaction is one half of 
that for the $\pi^+\pi^-\alpha$ when the two pions have equal momentum vectors.
With this prescription the interval of deuteron kinetic energies
$T_{\pk{d}}=787-1412$~MeV is mapped onto a proton range of the
\mbox{pp $\rightarrow \pk{d}\pi^+$} input $T_{p}=410-720$~MeV.

Since in the linear Fermi momentum approximation only the $\delta$-function 
survives the integration of eq.(\ref{e_1}), we define the form factor 
\begin{equation}
\label{e_4}
    \mathcal{W} = \frac{1}{\gamma^{2}\sqrt{\pk{m_{\alpha}\pi}}}
    \int \dkub{q_{1}}\dkub{q_{2}}\,
    \varphi_{\pk{d}}(\bmath{q}_{1}')
    \varphi_{\pk{d}}(\bmath{q}_{2}')
    \psi_{\alpha}^{\dagger}(\bmath{q}_{\alpha}) \:    
\delta\left(q_{1}^{z}+q_{2}^{z}+(\omega_1-\omega_{2})/2v_{\pk{d}}\right)\:.
\end{equation}

This integral is most easily evaluated by transforming into configuration 
space:
\[
    \mathcal{W} = \frac{2\pi}{\sqrt{\pk{m_{\alpha}}}}
        \int b\,\dd b\,\dd z_{1}\dd z_{2}\,
        \Phi_{\pk{d}}(\bmath{b},\gamma z_{1})\,
        \Phi_{\pk{d}}(-\bmath{b},\gamma z_{2})\,
        \Psi^{\dagger}_{\alpha}\left(\bmath{b},\frac{z_{1}-z_{2}}{2}\right)
       \times \]
\begin{equation}
\label{e_5}
    J_{0}\{|\bmath{k}_{1}^{b}-\bmath{k}_{2}^{b}|b/2\}
        \cos\{(k_{1}^{z}-k_{2}^{z})(z_{1}-z_{2})/4\}\,
      \cos\{(\omega_{1}-\omega_{2})(z_{1}+z_{2})/4v_{\pk{d}}\}\:.
\end{equation}

After averaging over the spin directions of the initial deuterons, the matrix 
element squared becomes 
\begin{equation}
\label{e_6}
    \frac{1}{9}\sum |\mathcal{M}|^{2} =
    \frac{\pk{m_{\alpha}^{2}}N_{\alpha}^{2}}
    {9v_{\pk{d}}^{2}}\, |\mathcal{W}|^{2}\,
    |\mathcal{C}|^{4}
    \left\{ 3(\ip{\bmath{\hat{p}}}{\bmath{k}_{1}})
    (\ip{\bmath{\hat{p}}}{\bmath{k}_{2}})
    +\ip{\bmath{k}_{1}}{\bmath{k}_{2}} \right\}^{2},
\end{equation}
where $N_{\alpha}^2$ is the number of deuteron pairs in the $\alpha$-particle.
In the ABC region where ${\bmath{k}_{1}}\approx {\bmath{k}_{2}}$, this
yields a characteristic $(3\cos^{2}\theta_{\alpha}+1)^{2}$ angular distribution,
which is just the square of that for the \mbox{pp $\rightarrow \pk{d}\pi^+$} 
differential cross section in the case of a pure $^{1\!}D_{2}p$ amplitude. This
is precisely what one would expect in a classical picture. On the other hand in 
the central bump the $\alpha$-particle distribution is isotropic and it is the 
pion distribution with respect to the beam direction which inherits this shape. 

The final expression for the differential cross section in the laboratory is
\begin{eqnarray}
    \left[ \frac{\dd^{2}\sigma}{\dd\Omega\, \dd p_{\alpha}} \right]_{\rm lab}
    & = &
    \frac{1}{192(2\pi)^{5}} \frac{p_{\alpha}^{2}k^{\ast}}
    {\pk{m_{d}}\,M_{\pk{X}}\,p_{\pk{d}}E_{\alpha}}
    \int \dd\Omega^{\ast} \sum |\mathcal{M}^{\ast}|^{2},
\label{e_7}
\end{eqnarray}
where for quantities denoted by an asterix $(^{\ast})$ the pion momenta and
angles are evaluated in the $\pi\pi$ rest frame. Though we have only calculated 
$\pi^+\pi^-$ production, from isospin arguments the corresponding cross section 
for neutral pions should be half of this. However, because of the narrowness of 
the ABC peak, it is important to evaluate the kinematics separately in the two 
cases.

In the numerical evaluation we use the $S$-state Paris wave 
function~\cite{paris} for the deuterons and the $\alpha$:dd cluster wave 
function of Forest et al.~\cite{dumbbell}. According to the latter, the 
$I=0$, $S=1$ $S$-state nucleon-nucleon distributions look very similar to those
of the deuteron when calculated with the corresponding nucleon-nucleon force 
provided that the separation $r_{\pk{np}}<2$ fm. The scale factor between the
two intensities is 4.7. Since pion absorption in the deuteron occurs primarily 
when the nucleons are close together, we need only the number of such `small' 
deuterons in the $\alpha$-particle. Thus, assuming that the `deuterons' in the 
$\alpha$-particle have independent distributions, we normalise the 
$\alpha$:dd wave function to $N_{\alpha}^2=2.3$.

\begin{figure}[p]
\begin{center}
    \includegraphics*[6mm,0mm][136mm,140mm]{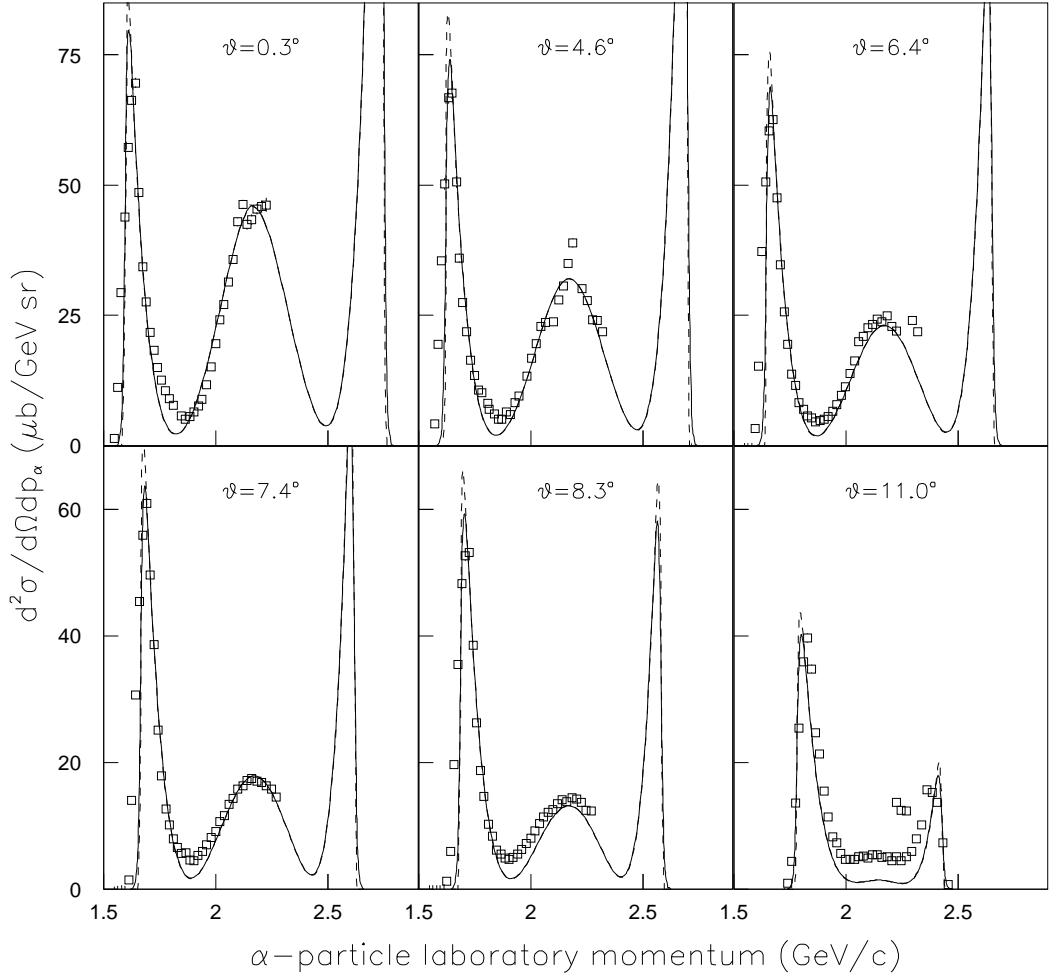}
\end{center}
\caption{The $\alpha$-particle momentum spectra from the 
\mbox{dd $\rightarrow\alpha$X} reaction for six different laboratory angles at
$T_{\rm d}=1250$~MeV. The experimental data of ref.~\protect\cite{Ban} are 
compared with the predictions of our double-pion-production model after
division by an overall factor of two. The broken lines are the raw calculations
and yield the solid lines after smearing over the experimental momentum
resolution.}
\label{fig:Ban1250}
\end{figure}

The only beam energy for which the Saclay group measured an angular distribution
is $T_{\pk{d}}=1250$~MeV \cite{Ban} and this corresponds to a region where the 
$^{1\!}D_{2}p$ transition \mbox{pp $\rightarrow \pk{d}\pi^+$} amplitude is
dominant~\cite{SAID}. When comparing with such $\alpha$-particle momentum 
spectra, it is important to take into account the momentum resolution and this
has been done by smearing the theoretical predictions over a resolution function
with $\sigma= 10$~MeV/c. After doing this, our predictions are typically a 
factor of two too high and, for ease of comparison, they have been divided by 
that in fig.~\ref{fig:Ban1250}. Such a small reduction in overall normalisation
could easily be due to initial distortion in the deuteron-deuteron system 
combined with uncertainty in the $\alpha$-particle normalisation.

The overall agreement between theory and experiment in fig.~\ref{fig:Ban1250} is
remarkably good and suggests strongly that, at least in this channel, the ABC 
is indeed a kinematic enhancement due to two independent $p$-wave pion 
productions. At a laboratory angle of $11^0$, the c.m.\ angle in the ABC peak is
about $45^0$ and so the predictions are following the change of about a factor
of 2.5 in c.m.\ differential cross section. The small deviations at the valleys
and the central bump at the larger angles may be due to having limited 
ourselves to just one amplitude. Away from 1250~MeV data the $^{1\!}D_{2}p$ 
dominance assumption is less good and, though we reproduce the other Saclay 
data \cite{Ban} to better than a factor of two, the details of the momentum 
spectra must wait until we have included a more complete set of 
\mbox{pp $\rightarrow \pk{d}\pi^+$} amplitudes in the calculation. This is 
currently in progress. 

A fuller set of input amplitudes is also required in order to predict the
deuteron analysing powers which have been measured near the forward direction 
in our energy domain at the SPESIII spectrometer at Saclay \cite{Nicole}. 
In particular the $\ip{\beps_{1}}{\beps_{2}}$ form in eq.(\ref{e_1})
predicts zero deuteron tensor analysing power whereas experiment yields a small
value but with significant structure which might arise through interferences
with other \mbox{pp $\rightarrow \pk{d}\pi^+$} amplitudes.

Extra confirmation of our approach might be provided by an exclusive measurement
of the \mbox{dd $\rightarrow\alpha\pi^+\pi^-$} reaction, since in the central
region we predict that the angular distribution should behave like
$(3\cos^2\theta_{\pi}+1)^2$.

We should like to thank Pia Th\"{o}rngren Engblom of the Stockholm Nuclear
Physics Group whose experiment at a much lower energy at the CELSIUS
ring~\cite{Pia} was the driving force behind the initiation of this study.
This work has been made possible by the continued financial support of the
Swedish Royal Academy and the Swedish Research Council, and one of the authors 
(CW) would like to thank them and the The Svedberg Laboratory for their 
generous hospitality.

%
%

\end{document}